\documentclass[sigconf,nonacm]{acmart}
\usepackage{enumitem}

\begin{document}

\title{Can Celebrities Burst Your Bubble?}

\author{Tu\u{g}rulcan Elmas}
\affiliation{%
  \institution{EPFL}
}
\email{tugrulcan.elmas@epfl.ch}

\author{Kristina Hardi}
\affiliation{%
  \institution{EPFL}
}
\email{kristina.satara@epfl.ch}

\author{Rebekah Overdorf}
\affiliation{%
  \institution{EPFL}
}
\email{rebekah.overdorf@epfl.ch}

\author{Karl Aberer}
\affiliation{%
  \institution{EPFL}
}
\email{karl.aberer@epfl.ch}

\begin{abstract}
  Polarization is a growing, global problem. As such, many social media based solutions have been proposed in order to reduce it. In this study, we propose a new solution that recommends topics to celebrities to encourage them to join a polarized debate and increase exposure to contrarian content --- bursting the filter bubble. Using a state-of-the art model that quantifies the degree of polarization, this paper makes a first attempt to empirically  answer the question: \emph{Can celebrities burst filter bubbles?} We use a case study to analyze how people react when celebrities are involved in a controversial topic and conclude with a list possible research directions.
\end{abstract}

\maketitle

\section{Introduction}

Polarization is a state in which the public is divided into groups with opposing opinions on an issue~\cite{definition}. Polarization is regarded as a threat to democracy and is detrimental to a healthy dialogue in a community. Echo chambers --- the phenomena in which individuals only hear the side of a debate they already agree with --- are a primary driver of polarization, as they are where extremist ideas foster~\cite{sunstein2001republic}. Social media platforms themselves hold some responsibility for the formation of such echo chambers; the algorithms that determine the information diet of users are believed to rank belief-reinforcing information higher as a result of maximizing engagement, which in turn minimizes cognitive dissonance.  The term \emph{filter bubble}, coined by Eli Pariser~\cite{pariser2011filter}, describes this phenomenon by which echo chambers are caused by the design of the system. 

Due to their potential detriment to democracy and society, others have proposed methods to burst filter bubbles in order to reduce polarization. Many of these solutions rely on action by the social media platforms itself, ignoring the fact that social networks have created this problem and may not be incentivised to act, including recommending users~\cite{garimella2017reducing} (aided by intermediary topics~\cite{graells2014people}) or content~\cite{lex2018mitigating} with opposing opinions, and presenting the information in a different way (i.e. showing the credibility of a source)~\cite{vydiswaran2015overcoming}. Other work focuses on raising awareness to users and therefore requires action on the part of the users who are in a filter bubble. These include exposing users to contrarian news~\cite{garimella2017mary}, raising awareness of one's connections' and own biases~\cite{gao2018burst}, convincing some users in the social network to reduce the overall polarization via education~\cite{matakos2017measuring}.

We present a new recommendation scheme that requires neither buy in from the social network nor action on the part of those in the filter bubble. This scheme bursts filter bubbles by recommending polarizing topics to influential users, i.e. celebrities. Prior work has shown that celebrities increase exposure and influence opinion in a controversial subject like vaccination~\cite{alatas2019celebrities}, implying that their involvement in a debate can reduce polarization by means of exposing users to counter opinions. Other work has shown that users value connections while selecting content~\cite{Messing2014SelectiveEI}, so messages conveyed by celebrities whom users are connected to, are likely to be valued over content from non-connections. Finally, this method leverages the fact the social media is a small-world network~\cite{small-world} so people with counter opinions are connected to the same users who do not explicitly posit their opinions. This is confirmed by ~\cite{hayat2019can} which shows that LeBron James is both followed by liberals and conservatives and also is the only liberal source in most of his conservative followers' profiles and therefore is able to burst those users' bubbles.

We identified the following research questions related to this scheme and answer them in the remainder of this paper:

\begin{enumerate}[label=\textbf{{RQ{\theenumi}}}]
    \item If a celebrity joins a debate on a controversial topic, will exposure to the contrarian content be increased and will polarization be reduced?
    \item How to select celebrities that would lead to such effects?
    \item Will such exposure mitigate the extreme opinions and hence decrease polarization?
    \item How would people react when a celebrity joins the debate? Will we observe mitigation of thoughts or backfire effect? 
\end{enumerate}


We define celebrity as ``anyone popular and although not strictly impartial, not politically polarized.'' In order to address \textbf{RQ1} and \textbf{RQ2}, we empirically show that inclusion of popular and neutral nodes into a polarized graph decreases the polarization of the graph, hence celebrities inclusion reduces polarization. To address \textbf{RQ3} and \textbf{RQ4}, we perform a qualitative analysis of users' reactions to celebrities participating in a controversial topic.

\section{Empirical Results on Effect of Celebrity Involvement}

\subsection{Theoretical Background}

First, we address \textbf{RQ1} to determine the effect that a celebrity joining a debate will have on polarization.
We use the quantifying controversy model~\cite{quantifying} which quantifies the controversy of a topic by computing how likely a user on one side of a polarized debate is to be exposed to content disseminated by a popular user on the opposing side. As such, this model serves as a proxy for polarization score on a topic. We recap the model briefly before detailing our application of it. 

\paragraph{Quantifying Controversy Model} First, consider a social graph $G(V,E)$ in which vertices $V$ are users who hold an opinion on a topic and edges $E$ are the social connections between them. $G$ is partitioned into two disjoint sets of users, $X$ and $Y$, which possibly correspond to the two different sides of the discussion. For each node in a set of randomly selected nodes, a random walk is started and concludes when it reaches any high-degree user. Let $P_{AB}$ be the probability that a random walk begins in partition $A$ and ends in partition $B$. The ``Random Walk Controversy Score'' (RWC) is the difference of the probabilities that a random walk begins and ends in the same partition ($P_{XX}$ and $P_{YY}$) and that a random walk begins and ends in the other partition ($P_{XY}$ and $P_{YX}$). 

$$RWC = P_{XX}P_{YY} - P_{XY} P_{YX}$$

Resulting $RWC$ score is inversely correlated with likeliness of exposure to popular content from opposite side and implies polarization of the debate on the topic. 


Since the users in the same partition are well connected due to homophily principle, we assume that in a polarized network content produced in the same partition has the same stance. Conversely, content produced by users in different partitions have different stances. Thus, we can claim that in a polarized network, content that is more quickly reached by a user (via a random walk) is from the same stance as the user. Hence, the user is trapped in an echo chamber. We leave a model for echo chambers which works without this assumption about stances to future work. 


In the context of Twitter, the topic is modeled as tweets containing relevant hashtags to a seed hashtag that defines the topic. The social network, $G$, is built by including users who authored these tweets. The links between the nodes of these networks could be following, retweeting or both. We use following relationships as they are better proxy to measure exposure to content from the connected user. The users who are recommended the topic will be added to the $G$ if they accept the recommendation. Links between users already in $G$ and a newly added user will be added to $G$ if these users already follow the newly added user. Our problem is then to identify such nodes to recommend the topic so that their inclusion in the network will decrease $RWC$ of $G$.

One issue with this topic modeling approach is that it draws mostly from politically interested users and hence exaggerates the polarization of a popular topic that involves many hashtags and keywords~\cite{tufekci2014big}. However, this approach is plausible when you consider the scenario in which a user clicks on a hashtag about a controversial topic that is trending. We assume that the tweets from users' connections are more likely to be ranked higher and hence the user will be in a filter bubble when presented with tweets on that topic.

\subsection{Node Addition Problem}

In order to determine which celebrities to select (\textbf{RQ2}), we define the Node Addition Problem as determining which nodes to add to the network in order to maximize the reduction of the controversy of the topic. Consider a topic $T$ and a social graph \mbox{$G = (V, E)$} made up of users who participated in a debate about $T$. Let the controversy score of $G$ be $RWC(G)$. Let the \emph{Potential Social Graph} $G^{**} (V^{**}, E^{**})$ be the union of \mbox{$G = (V, E)$} and all the vertices that are connected to $V$ but did not discuss $T$ and the edges connecting them to $V$. The node addition problem is to find a set of $k$ nodes $V'$ not in $G$ but in $G^{**}$ to add to $G$ and obtain the \emph{Augmented Graph} $G^{*} = (V^{*}, E^{*})$ which maximizes $RWC(G)$ - $RWC(G{*})$.

We hypothesize that $k$ nodes that maximize the decrease of $RWC$ will be those who have 1) high in-degree 2) edges distributed evenly between two partitions. We leave mathematical proof for future work and only present empirical results. To find those $k$ nodes, we use the Fagin algorithm \cite{fagin2003optimal} to rank nodes by 1) their in-degree and 2) their minimum ratio of connections' to one partition over all connections. The first is a proxy for popularity and the second a proxy for neutrality. We compute $RWC$ for each candidate and choose the $k$ nodes which yield the largest $RWC$ decrease. We assume these $k$ nodes will consist of candidates which minimize $RWC$ individually to avoid computing $RWC$ for every $k$ combination of nodes, which is very costly. 

\subsection{Experimental Results}

\begin{figure}
\graphicspath{{russiamarch.png}}
\includegraphics[width=1\linewidth]{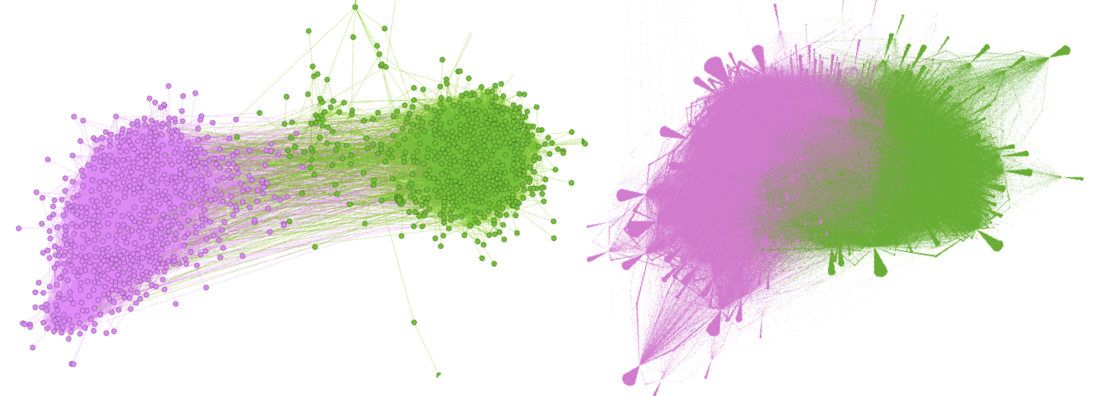}
\caption{Social Graph (left) and Potential Social Graph (right). The nodes are users and the edges indicate follows. The colors indicate partitions. Force Directed Layout is used to visualize graphs. Notice that Potential Social Graph is not polarized like Social Graph.}
\label{fig:russiamarch}
\end{figure}

For the empirical results, we used the follower data of users who participated in the debate about the topic \#Russia\_March. This topic is studied in~\cite{quantifying} and is already found to be polarized. We first created the follower graph $G$ which involves only the users who tweeted with \#Russia\_March and relevant hashtags. Then we collected all the followees of users in the $G$ to create the augmented graph. See Figure \ref{fig:russiamarch} for the two graphs.

\begin{figure}
\graphicspath{{addnodesdecreaserwc.png}}
\includegraphics[width=1\linewidth]{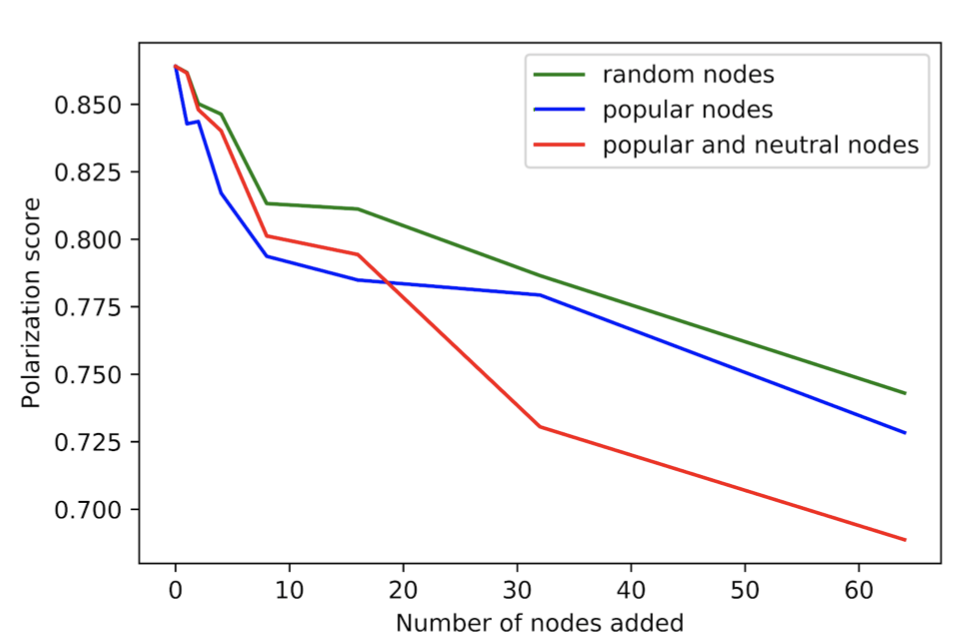}
\caption{Decrease in polarization score according to number of nodes added. Colors denote the method to choose the nodes to be added. Notice that choosing popular and neutral nodes to recommend topic is much more effective than merely choosing popular nodes after 20 additions. }
\label{fig:experiment}
\end{figure}

We used two baselines to evaluate our node selection process. First, the most popular nodes to study the effect neutrality and second  ``random nodes'', which are artificially created nodes that have fixed degree (50) and are connected to 25 randomly chosen nodes in each partition to study the effect of popularity. Figure \ref{fig:experiment} shows the final polarization score with respect to the number of nodes added. Colors denote the method used to choose nodes. Although adding popular nodes seems beneficial initially, they become ineffective after 20 nodes. As the results indicate, most popular and neutral nodes reduce polarization, who happen to be celebrities by our definition. 

\begin{figure}
\graphicspath{{losefollowers.png}}
\includegraphics[width=1\linewidth]{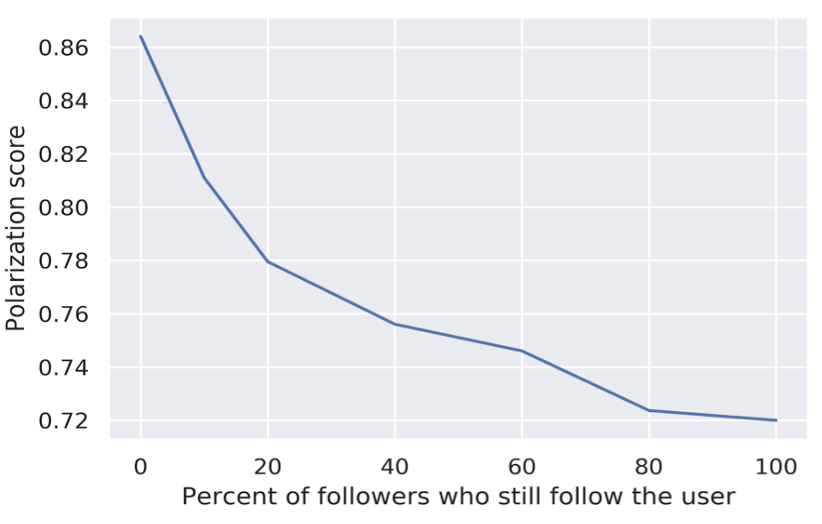}
\caption{Simulation of a scenario where users unfollow the celebrity who joined the debate. The polarization is still reduced effectively even when 20\% of the users still follow the celebrity.}
\label{fig:losefollowers}
\end{figure}

A possible side effect of this method is that users will unfollow celebrities who discuss controversial topics and join the debate. To simulate this, we randomly break incoming links of the $k$ nodes. As seen in Figure \ref{fig:losefollowers}, the polarization is still reduced unless the celebrities lost most of their followers, which is unrealistic. 

The empirical results show that polarization as measured by $RWC$ reduces when celebrities join a controversial debate. It is not clear, however, how much reduction in $RWC$ equates to real life implications. In the next section, we explore what happens in a real life study of celebrities weighing in on a controversial topic. 

\section{When Celebrities Break Their Silence: Observations from the Case of 2019 \.{I}stanbul Election Rerun Decision}

The 2019 \.{I}stanbul Election Rerun was a controversial decision by the Supreme Electoral Council when the opposition's candidate won over ruling AKP's candidate by slim margin despite high voter turnout. 
The decision was deemed unfair by supporters of the opposition. Many celebrities started to tweet after the opposition candidate, Ekrem \.{I}mamo\u{g}lu, gave a speech and said ``Everyone should speak, the celebrities should speak!''~\cite{herkeskonusacak}. This serves as a suitable case study as those celebrities' messages reached a very wide audience; many have more than 100,000 followers. 

We address \textbf{RQ4} and determine how users react to celebrities joining a debate by analyzing immediate reactions to the celebrity tweets regarding this decision. We also address \textbf{RQ3} on this real world data set and determine via a longitudinal study if this celebrity intervention really made a difference in this case.


We selected 81 celebrities from a list of Turkish celebrities~\cite{eksilist} who tweeted in favor of the opposition's candidate on the night of 6 May 2019. We collected their tweets, retweets, and replies to their tweets for one week using Twitter's Streaming API. By manual inspection, we found that 47 of them are in cinema-tv business, 24 in music, and 10 in other fields. Judging from the tweets since September 2019, 43 celebrities were already found to be posting frequently about controversial but apolitical topics and 7 were found to be occasionally posting about such topics. The remaining 31 only post personal and professional updates. No celebrity showed explicit political affiliation or criticism towards a party or government, but 25 criticized recent government policies, 15 infrequently posted content that could be interpreted as anti-government, and the remaining 41 appear politically neutral. This suggests that for most of the celebrities in our dataset, the \.{I}stanbul Election Rerun was the first time they spoke out on a political topic on Twitter.

For those celebrities with more than one relevant tweet, we selected the tweet that received the most replies for each celebrity in 6-7 May 2019, then annotated each celebrity according to their stance with respect to that tweet. Note that 60 of the celebrities showed explicit support to opposition's candidate (30 used the opposition slogan \#Her\c{S}ey\c{C}okG\"{u}zelOlacak), while 8 celebrities only commented on the unfairness of the decision of rerun. In addition, 11 celebrities called for citizens to vote in the rerun, and 2 called for other celebrities to tweet. 



We randomly sampled 10 direct replies per celebrity tweet. We annotated these replies according to 1) stance on the celebrity (positive, negative, neutral) and 2) narratives they contain. Note that not all  replies had a narrative. We removed the celebrity tweets that were irrelevant or had less than 10 replies. We annotated 679 tweets in total. We found that 434 tweets had a positive stance towards the celebrity and their idea, 178 tweets had a negative stance (with 31 containing insults), and 60 tweets had a neutral stance. Our analysis indicates the following narratives are prevalent in tweets, which have a non-positive stance unless otherwise specified.

\begin{enumerate}
    \item Counter argument: The celebrity is wrong as the opposition has committed voter fraud and the decision was correct. (n = 29) 
    
    \item Ad hominem: The celebrity is wrong or does not deserve a voice on the matter due to their past political actions, or their character is not harmonious with their idea. (n = 26) 
    
    \item Self interests: The celebrity is behaving this way not because of patriotism but for self interest because they are not successful in their work. (n = 19)
    
    \item Questioning authority: The celebrity does not have a right to speak because they are not a real celebrity. (n = 16) 
    
    \item Whataboutism: The celebrity's patriotism is in question as they did not react to soldiers killed by terrorist attacks or on the night of 15 July 2016 coup. (n = 11)
    
    \item Too late: Positive with the celebrity's opinion but blames them or celebrities in general for acting too late. (n = 10)
    
    \item Reactionary: The celebrities should not expose their political beliefs or champion one political side or should be remembered with their art only. (n = 5)
    

    \item Hopeless: The situation is hopeless and they will not win the rerun, although the celebrity is championing hope. (negative: n = 3), (positive: n = 2) 
    
    
    \item Mitigate: Indicates a non-polarized affiliation, but agrees with the celebrity on that issue with positive stance. (n = 2) 
    
    \item Backfire: Threatens the celebrity (n = 9), 
    or indicates they will no longer follow them (n = 3).
    
\end{enumerate}

Based on this analysis, we make the following observations:

\textbf{The celebrities' messages reached users with the opposite stance:} The presence of negative replies from the opposite side shows that the goal of bursting the filter bubbles has been achieved.

\textbf{The source of the message matters:} if the celebrity is not having a successful career and is not respected, they have negative reactions implying those elements. Even the supportive replies had sarcastic elements in some cases. Their past political deeds also matter especially if they took a pro-government stance before. 

\textbf{The content of the message also matters:} if the content is sarcastic or has some logical flaw, the replies indicate this rather than agreement or disagreement which causes distraction. The reactions which would lead to meaningful discussions (although still rare) come when the celebrity's tweet contains an argument. 

\textbf{There is no evidence of correlation between political activity on Twitter and stances of replies a celebrity gets:} We averaged the stances of replies each celebrity gets (1 for positive, 0 for neutral and -1 for negative.) and ran a t-test for those who were political / commented on contemporary issues on Twitter and those who do not. The stance of replies turned out to be independent of the both factors as the p-value was insignificant. 

\textbf{The negative replies mainly come from politically motivated users:} We inspected 100 users who had a negative stance and replied to celebrities. Three of these users had deleted their accounts and 15 were suspended. Among the 82 remaining users, 55 were very polarized --- their account seemed to be opened only to share pro-AKP content, and they constantly spread fake news about the opposition. This suggest that the replies should not be taken as genuine public reaction during analysis as they also likely part of coordinated attacks. However the narratives they contain are still important as they may influence genuine Twitter users. 

\textbf{Both mitigation and backfire effect appear to be small:} Inferring from the reactions, we had only two cases where an artist's presence made a positive effect. Backfire effect is also small; follower counts increased rather than decreased, which may show that they do not go out of favour dramatically.

\section{Open Questions}

\hspace{10pt}\textbf{Celebrity acceptance:} Would a celebrity accept the recommendation to join the debate 1) by public request, 2) by platform request, 3) by a fellow celebrity, or 4) not at all?
    
\textbf{Factors on people's reactions:} Are people's reactions to celebrities joining to a political debate dependant on the side they join, on whether they try to mitigate extreme opinions or not, or whether multiple celebrities observe the same behavior?
    
\textbf{Modification of the platform:} What would be the effects of a modification to the platform so that 1) it recommends topic to users who would increase exposure to the contrarian content and decrease polarization and 2) it recommends content by such users to users with extreme views to be mitigated?

\textbf{Categorization of celebrity candidates:} Not all popular accounts are suitable to recommend topics to comment on, i.e. corporate and media related accounts will be unlikely to take the recommendation for fear of backlash.

\textbf{When do celebrities pick up on a controversial topic?} Do they pick up on early and help the topic spread, or they pick up on late? If late, is it because peer pressure, for self interests, or neither? Such analysis would be helpful to determine if recommending topics to celebrities is realistic or helpful. 

\textbf{Non-political users:} If many celebrities are tweeting about political topics due to this method, users who use Twitter for entertainment purposes and not for political engagement are negatively impacted and may leave the platform.

\textbf{Revision of quantifying polarization algorithms:} Current algorithms do not scale, do not take temporal signals into account, and do not take graph modularity due to external factors like language into account.

\textbf{RT = Endorsement?} Quantifying polarization studies assume that social connections like retweets and follows are endorsements without justification. However in real life, this assumption falls apart in many cases. Many users even \emph{explicitly state} that retweets are not endorsements on their profiles. In some cases, endorsement ocurrs without a like: videos involving \#BLM (Black Lives Matter) protests and police intervention on Facebook were not liked but shared~\cite{tufekci2017twitter}. In Twitter terms, this would mean that newsworthy posts with a negative sentiment are not liked but retweeted, which breaks this assumption. A survey among 316 users revealed that only 68\% of people endorse what they retweet, and 73\% of users agree with what they retweet~\cite{metaxas2015retweets}. More work to come up with better connection models is needed for work on polarization to remain legitimate.

\textbf{Revision of identification of a topic:} Hashtags do not capture all the discussion on a topic and focus attention on already polarized users, thus creating biased results. Therefore, methodology to model a topic should be revised.

\textbf{Backfiring effect:} Anti-polarization tools assume that views will be moderated when a user is connected to users of opposite view by the implicit assumption that views will be averaged, ignoring the possibly backfire effect.

\textbf{Universal Interest:} There is an underlying assumption that an unbiased user has a medium opinion. Most works do not consider that a user may have no opinion on a topic. 

\textbf{Lurkers  Matter:} The observations from Twitter analysis are based only on the audience that actively reacts. However Facebook users were found to underestimate their audience by 27\%~\cite{bernstein2013quantifying}. We expect a more dramatic result on Twitter since most profiles are public and timelines are created based on more than simple follow relationships. Thus, future work is needed to verify these results. 

\textbf{Not all views should be moderated:}. In fact it can be harmful in some cases to encourage users towards some position. For example, encouraging normal users to read anti-vac content could be detrimental to public health. 

\begin{acks}
This work is supported in part by the Open Technology Fund's Information Controls Fellowship.
\end{acks}

\bibliographystyle{ACM-Reference-Format}
\bibliography{citations}

\end{document}